\documentclass{article}%
\usepackage{amsfonts}
\usepackage{amsmath}

\providecommand{\U}[1]{\protect \rule{.1in}{.1in}}

 \evensidemargin0in \oddsidemargin0in \topmargin10pt
\begin{document}

\title{Wigner operator's new transformation in phase space quantum mechanics and its
applications \thanks{{\small Work supported by the National Natural Science
Foundation of China under grant: 10775097, 10874174, and Specialized research
fund for the doctoral program of higher education of China}} }
\author{$^{1,2}$Hong-yi Fan\\$^{1}${\small Department of Physics, Shanghai Jiao Tong University,
\ Shanghai, 200030, China}\\{\small \ }$^{2}${\small Department of Material Science and Engineering,}\\{\small \ University of Science and Technology of China, Hefei, Anhui 230026,
China}}
\maketitle

\begin{abstract}
Using operators' Weyl ordering expansion formula (Hong-yi Fan,\emph{\ }J.
Phys. A 25 (1992) 3443) we find new two-fold integration transformation about
the Wigner operator $\Delta \left(  q^{\prime},p^{\prime}\right)  $ ($q
$-number transform) in phase space quantum mechanics,
\[
\iint_{-\infty}^{\infty}\frac{\mathtt{d}p^{\prime}\mathtt{d}q^{\prime}}{\pi
}\Delta \left(  q^{\prime},p^{\prime}\right)  e^{-2i\left(  p-p^{\prime
}\right)  \left(  q-q^{\prime}\right)  }=\delta \left(  p-P\right)
\delta \left(  q-Q\right)  ,
\]
and its inverse%
\[
\iint_{-\infty}^{\infty}\mathtt{d}q\mathtt{d}p\delta \left(  p-P\right)
\delta \left(  q-Q\right)  e^{2i\left(  p-p^{\prime}\right)  \left(
q-q^{\prime}\right)  }=\Delta \left(  q^{\prime},p^{\prime}\right)  ,
\]
where $Q,$ $P$ are the coordinate and momentum operators, respectively. We
apply it to studying mutual converting formulas among $Q-P$ ordering, $P-Q$
ordering and Weyl ordering of operators. In this way, the contents of phase
space quantum mechanics can be enriched.

PACS: 03.65.-w, 02.90.+p

Keywords: Wigner operator; Weyl ordering; two-fold integration transformation

\end{abstract}

\section{Introduction}

Phase space quantum mechanics (PSQM) pioneered by Wigner [1] and Weyl [2] has
been paid more and more attention since the foundation of quantum mechanics,
because it has wide applications in quantum statistics, quantum optics, and
quantum chemistry. In PSQM observables and states are replaced by functions on
classical phase space so that expected values are calculated, as in classical
statistical physics, by averaging over the phase space. The phase-space
approaches provides valuable physical insight and allows us to describe alike
classical and quantum processes using the similar language. Development of
phase space quantum mechanics [3-5] always accompanies with solving operator
ordering problems. Weyl proposed a scheme for quantizing classical coordinate
and momentum quantity $q^{m}p^{n}$ ($c$-number) as the quantum operators
($q$-number) in the following way
\begin{equation}
q^{m}p^{n}\rightarrow \left(  \frac{1}{2}\right)  ^{m}\sum_{l=0}^{m}\binom
{m}{l}Q^{m-l}P^{n}Q^{l},\label{1}%
\end{equation}
where $Q,$ $P$ are the coordinate and momentum operators, respectively,
$[Q,P]=\mathtt{i}\hbar.$ (Later in this work we set $\hbar=1).$ The right-hand
side of (\ref{1}) is in Weyl ordering, so we introduced the symbol $%
\genfrac{}{}{0pt}{}{:}{:}%
\genfrac{}{}{0pt}{}{:}{:}%
$ to characterize it [6-7], and%
\begin{align}
q^{m}p^{n}  & \rightarrow \left(  \frac{1}{2}\right)  ^{m}\sum_{l=0}^{m}%
\binom{m}{l}Q^{m-l}P^{n}Q^{l}\nonumber \\
& =%
\genfrac{}{}{0pt}{}{:}{:}%
\left(  \frac{1}{2}\right)  ^{m}\sum_{l=0}^{m}\binom{m}{l}Q^{m-l}P^{n}Q^{l}%
\genfrac{}{}{0pt}{}{:}{:}%
=%
\genfrac{}{}{0pt}{}{:}{:}%
Q^{m}P^{n}%
\genfrac{}{}{0pt}{}{:}{:}%
,\label{2}%
\end{align}
where in the second step we have used the property that Bose operators are
commutative within $%
\genfrac{}{}{0pt}{}{:}{:}%
\genfrac{}{}{0pt}{}{:}{:}%
.$ This is like the fact that Bose operators are commutative within the normal
ordering symbol $:$ $:$. The Weyl quantization rule between an operator
$H\left(  P,Q\right)  $ and its classical correspondence is
\begin{equation}
H\left(  P,Q\right)  =\iint_{-\infty}^{\infty}\mathtt{d}q\mathtt{d}ph\left(
p,q\right)  \Delta \left(  q,p\right)  ,\label{3}%
\end{equation}
where $\Delta \left(  q,p\right)  $ is the Wigner operator [2-5] [8]. Using $%
\genfrac{}{}{0pt}{}{:}{:}%
\genfrac{}{}{0pt}{}{:}{:}%
$ we have invented the integration technique within Weyl ordered product of
operators with which we constructed an operators' Weyl ordering expansion
formula (see Eq. (21) below), which is the same as Eq. (53) in Ref. [6]). In
this work we shall use this formula to find new two-fold $q$-number
integration transformation about the Wigner operator $\Delta \left(  q^{\prime
},p^{\prime}\right)  $ in phase space quantum mechanics (see Eqs. (33) and
(34) below), which helps to convert P-Q ordering and Q-P ordering to Weyl
ordering, and vice versa. The work is arranged as follows: In Sec. 2 we
briefly review the Weyl ordered form of Wigner operator. In Sec. 3 we derive
the Weyl ordering forms of $\delta \left(  p-P\right)  \delta \left(
q-Q\right)  $ and $\delta \left(  q-Q\right)  \delta \left(  p-P\right)  ,$
their transformation to the Wigner operator is shown in Sec. 4. Based on Sec.
4 we in Sec. 5 propose a new $c$-number integration transformation in $p-q$
phase space, see Eq. (35) below, and its inverse transformation, which
possesses Parsval-like theorem. Secs. 6-8 are devoted to deriving mutual
converting formulas among $Q-P$ ordering, $P-Q$ ordering and Weyl ordering of
operators. In this way, the contents of phase space quantum mechanics can be enriched.

\section{The Weyl ordered form of Wigner operator}

According to Eq. (3) we can rewrite Eq. (2) as
\begin{equation}%
\genfrac{}{}{0pt}{}{:}{:}%
Q^{m}P^{n}%
\genfrac{}{}{0pt}{}{:}{:}%
=\iint \mathtt{d}q\mathtt{d}pq^{m}p^{n}\Delta \left(  q,p\right)  ,\label{4}%
\end{equation}
which implies that the integration kernel (the Wigner operator) is [6-7]
\begin{equation}
\Delta \left(  q,p\right)  =%
\genfrac{}{}{0pt}{}{:}{:}%
\delta \left(  q-Q\right)  \delta \left(  p-P\right)
\genfrac{}{}{0pt}{}{:}{:}%
=%
\genfrac{}{}{0pt}{}{:}{:}%
\delta \left(  p-P\right)  \delta \left(  q-Q\right)
\genfrac{}{}{0pt}{}{:}{:}%
.\label{5}%
\end{equation}
Substituting (5) into (3) yields $H\left(  P,Q\right)  =%
\genfrac{}{}{0pt}{}{:}{:}%
h\left(  P,Q\right)
\genfrac{}{}{0pt}{}{:}{:}%
,$ where $%
\genfrac{}{}{0pt}{}{:}{:}%
h\left(  P,Q\right)
\genfrac{}{}{0pt}{}{:}{:}%
$ is just the result of replacing $p\rightarrow P,q\rightarrow Q$ in $h\left(
p,q\right)  $ and then putting it within $%
\genfrac{}{}{0pt}{}{:}{:}%
\genfrac{}{}{0pt}{}{:}{:}%
.$ Further, using
\begin{equation}
Q=\frac{a+a^{\dagger}}{\sqrt{2}},\text{ \ }P=\frac{a-a^{\dagger}}{\sqrt
{2}\mathtt{i}},\text{ }\alpha=\frac{q+\mathtt{i}p}{\sqrt{2}},\text{ }\left[
a,a^{\dagger}\right]  =1,\label{6}%
\end{equation}
we can express%
\begin{equation}
\Delta \left(  q,p\right)  \rightarrow \Delta \left(  \alpha,\alpha^{\ast
}\right)  =\frac{1}{2}%
\genfrac{}{}{0pt}{}{:}{:}%
\delta \left(  \alpha-a\right)  \delta \left(  \alpha^{\ast}-a^{\dagger}\right)
%
\genfrac{}{}{0pt}{}{:}{:}%
.\label{7}%
\end{equation}
It then follows%
\begin{align}%
\genfrac{}{}{0pt}{}{:}{:}%
K\left(  a^{\dagger},a\right)
\genfrac{}{}{0pt}{}{:}{:}
& =\int \mathtt{d}^{2}\alpha K\left(  \alpha^{\ast},\alpha \right)
\genfrac{}{}{0pt}{}{:}{:}%
\delta \left(  \alpha-a\right)  \delta \left(  \alpha^{\ast}-a^{\dagger}\right)
%
\genfrac{}{}{0pt}{}{:}{:}%
\nonumber \\
& =2\int \mathtt{d}^{2}\alpha K\left(  \alpha^{\ast},\alpha \right)
\Delta \left(  \alpha,\alpha^{\ast}\right)  ,\label{8}%
\end{align}
Thus the neat expression of $\Delta \left(  q,p\right)  $ in Dirac's delta
function form is very useful, one of its uses is that the marginal
distributions of Wigner operator can be clearly shown, due to the coordinate
and momentum projectors are respectively
\begin{equation}
\left \vert q\right \rangle \left \langle q\right \vert =\delta \left(  q-Q\right)
=%
\genfrac{}{}{0pt}{}{:}{:}%
\delta \left(  q-Q\right)
\genfrac{}{}{0pt}{}{:}{:}%
,\label{9}%
\end{equation}%
\begin{equation}
\left \vert p\right \rangle \left \langle p\right \vert =\delta \left(  p-P\right)
=%
\genfrac{}{}{0pt}{}{:}{:}%
\delta \left(  p-P\right)
\genfrac{}{}{0pt}{}{:}{:}%
,\label{10}%
\end{equation}
we immediately know that the following marginal integration
\begin{equation}
\int_{-\infty}^{\infty}\mathtt{d}q\Delta \left(  q,p\right)  =\int_{-\infty
}^{\infty}\mathtt{d}q%
\genfrac{}{}{0pt}{}{:}{:}%
\delta \left(  q-Q\right)  \delta \left(  p-P\right)
\genfrac{}{}{0pt}{}{:}{:}%
=%
\genfrac{}{}{0pt}{}{:}{:}%
\delta \left(  p-P\right)
\genfrac{}{}{0pt}{}{:}{:}%
=\left \vert p\right \rangle \left \langle p\right \vert ,\label{11}%
\end{equation}
similarly,%
\begin{equation}
\int_{-\infty}^{\infty}\mathtt{d}p\Delta \left(  q,p\right)  =%
\genfrac{}{}{0pt}{}{:}{:}%
\delta \left(  q-Q\right)
\genfrac{}{}{0pt}{}{:}{:}%
=\left \vert q\right \rangle \left \langle q\right \vert .\label{12}%
\end{equation}
It then follows the completeness of $\Delta \left(  q,p\right)  ,$%
\begin{equation}
\iint \limits_{-\infty}^{\infty}\mathtt{d}q\mathtt{d}p\Delta \left(  q,p\right)
=1,\label{13}%
\end{equation}
so the Weyl rule for $H\left(  P,Q\right)  $ in (3) can also be viewed as
$H$'s expansion in terms of $\Delta \left(  q,p\right)  .$ When $H\left(
P,Q\right)  $ is in Weyl ordered, which means $H\left(  P,Q\right)  =%
\genfrac{}{}{0pt}{}{:}{:}%
H\left(  P,Q\right)
\genfrac{}{}{0pt}{}{:}{:}%
,$ then using the completeness (13) we see%
\begin{equation}%
\genfrac{}{}{0pt}{}{:}{:}%
H\left(  P,Q\right)
\genfrac{}{}{0pt}{}{:}{:}%
=%
\genfrac{}{}{0pt}{}{:}{:}%
H\left(  P,Q\right)
\genfrac{}{}{0pt}{}{:}{:}%
\iint \limits_{-\infty}^{\infty}\mathtt{d}q\mathtt{d}p\Delta \left(  q,p\right)
=\iint \limits_{-\infty}^{\infty}\mathtt{d}q\mathtt{d}pH\left(  q,p\right)
\Delta \left(  q,p\right)  ,\label{14}%
\end{equation}
as if $\Delta \left(  q,p\right)  $ was the "eigenvector" of $%
\genfrac{}{}{0pt}{}{:}{:}%
H\left(  P,Q\right)
\genfrac{}{}{0pt}{}{:}{:}%
.$ On the other hand, due to the normally ordered forms of $\left \vert
q\right \rangle \left \langle q\right \vert $ and $\left \vert p\right \rangle
\left \langle p\right \vert $ [8]
\begin{equation}
\left \vert q\right \rangle \left \langle q\right \vert =\frac{1}{\sqrt{\pi}%
}\colon e^{-\left(  q-Q\right)  ^{2}}\colon,\label{15}%
\end{equation}%
\begin{equation}
\left \vert p\right \rangle \left \langle p\right \vert =\frac{1}{\sqrt{\pi}%
}\colon e^{-\left(  p-P\right)  ^{2}}\colon,\label{16}%
\end{equation}
we know the normally ordered form of $\Delta \left(  q,p\right)  $ [9]
\begin{equation}
\Delta \left(  q,p\right)  =\frac{1}{\pi}\colon e^{-\left(  q-Q\right)
^{2}-\left(  p-P\right)  ^{2}}\colon=\frac{1}{\pi}\colon e^{-2\left(
\alpha^{\ast}-a^{\dagger}\right)  \left(  \alpha-a\right)  }\colon
=\Delta \left(  \alpha,\alpha^{\ast}\right)  .\label{17}%
\end{equation}
Using the completeness relation of the coherent state $\left \vert
\beta \right \rangle ,$
\begin{equation}
\int \frac{d^{2}\beta}{\pi}\left \vert \beta \right \rangle \left \langle
\beta \right \vert =1,\text{\ }\left \vert \beta \right \rangle =\exp[-\frac
{|\beta|^{2}}{2}+\beta a^{\dagger}]\left \vert 0\right \rangle ,\text{
\ }a\left \vert \beta \right \rangle =\beta \left \vert \beta \right \rangle
,\label{18}%
\end{equation}
where $\left[  a,a^{\dagger}\right]  =1,$ $\left \vert \beta \right \rangle $ is
the coherent state [10-11], we have%
\begin{align}
2\pi \mathtt{Tr}\Delta \left(  \alpha,\alpha^{\ast}\right)   & =2\mathtt{Tr}%
\left[  \colon e^{-2\left(  \alpha^{\ast}-a^{\dagger}\right)  \left(
\alpha-a\right)  }\colon \int \frac{\mathtt{d}^{2}\beta}{\pi}\left \vert
\beta \right \rangle \left \langle \beta \right \vert \right] \nonumber \\
& =2\int \frac{\mathtt{d}^{2}\beta}{\pi}e^{-2\left(  \alpha^{\ast}-\beta^{\ast
}\right)  \left(  \alpha-\beta \right)  }=1,\label{19}%
\end{align}
this is equivalent to (13). Using (17) we also easily obtain
\begin{align}
& \mathtt{Tr}\left[  \Delta \left(  \alpha,\alpha^{\ast}\right)  \Delta \left(
\alpha^{\prime},\alpha^{\prime \ast}\right)  \right] \nonumber \\
& =\frac{1}{\pi^{2}}\mathtt{Tr}\left[  \colon e^{-2\left(  \alpha^{\ast
}-a^{\dagger}\right)  \left(  \alpha-a\right)  }\colon \int \frac{\mathtt{d}%
^{2}\beta}{\pi}\left \vert \beta \right \rangle \left \langle \beta \right \vert
\colon e^{-2\left(  \alpha^{\prime \ast}-a^{\dagger}\right)  \left(
\alpha^{\prime}-a\right)  }\colon \right] \nonumber \\
& =\mathtt{Tr}\left[  \int \frac{\mathtt{d}^{2}\beta}{\pi^{3}}e^{-2\left(
\alpha^{\ast}-a^{\dagger}\right)  \left(  \alpha-\beta \right)  }\left \vert
\beta \right \rangle \left \langle \beta \right \vert e^{-2\left(  \alpha
^{\prime \ast}-\beta^{\ast}\right)  \left(  \alpha^{\prime}-a\right)  }\right]
\nonumber \\
& =\int \frac{\mathtt{d}^{2}\beta}{\pi}\left \langle \beta \right \vert
e^{-2\left(  \alpha^{\prime \ast}-\beta^{\ast}\right)  \left(  \alpha^{\prime
}-a\right)  }e^{-2\left(  \alpha^{\ast}-a^{\dagger}\right)  \left(
\alpha-\beta \right)  }\left \vert \beta \right \rangle \nonumber \\
& =\int \frac{\mathtt{d}^{2}\beta}{\pi}e^{-2\left(  \alpha^{\ast}-\beta^{\ast
}\right)  \left(  \alpha-\beta \right)  -2\left(  \alpha^{\prime \ast}%
-\beta^{\ast}\right)  \left(  \alpha^{\prime}-\beta \right)  }e^{4\left(
\alpha-\beta \right)  \left(  \alpha^{\prime \ast}-\beta^{\ast}\right)
}\nonumber \\
& =\int \frac{\mathtt{d}^{2}\beta}{\pi^{3}}e^{2\beta^{\ast}\left(
\alpha^{\prime}-\alpha \right)  -2\beta \left(  \alpha^{\prime \ast}-\alpha
^{\ast}\right)  -2|\alpha|^{2}-2|\alpha^{\prime}|^{2}+4\alpha \alpha
^{\prime \ast}}\nonumber \\
& =\frac{1}{4\pi}\delta \left(  \alpha-\alpha^{\prime}\right)  \delta \left(
\alpha^{\ast}-\alpha^{\prime \ast}\right)  .\label{20}%
\end{align}

\section{Weyl ordering of $\delta \left(  p-P\right)  \delta \left(  q-Q\right)
$ and $\delta \left(  q-Q\right)  \delta \left(  p-P\right)  $}

In Refs. [6-7] we have presented operators' Weyl ordering expansion formula%
\begin{equation}
\rho=2\int \frac{\mathtt{d}^{2}\beta}{\pi}%
\genfrac{}{}{0pt}{}{:}{:}%
\left \langle -\beta \right \vert \rho \left \vert \beta \right \rangle \exp \left[
2\left(  \beta^{\ast}a-a^{\dagger}\beta+a^{\dagger}a\right)  \right]
\genfrac{}{}{0pt}{}{:}{:}%
.\label{21}%
\end{equation}
For the pure coherent state density operator $\left \vert \alpha \right \rangle
\left \langle \alpha \right \vert ,$ using (21) and the overlap $\left \langle
\alpha \right \vert \left.  \beta \right \rangle =\exp[-\frac{1}{2}\left(
|\alpha|^{2}+|\beta|^{2}\right)  +\alpha^{\ast}\beta]$ we derive
\begin{align}
\left \vert \alpha \right \rangle \left \langle \alpha \right \vert  & =2%
\genfrac{}{}{0pt}{}{:}{:}%
\int \frac{\mathtt{d}^{2}\beta}{\pi}\left \langle -\beta \right \vert \left.
\alpha \right \rangle \left \langle \alpha \right \vert \left.  \beta \right \rangle
\exp[2\left(  \beta^{\ast}a-a^{\dagger}\beta+a^{\dagger}a\right)  ]%
\genfrac{}{}{0pt}{}{:}{:}%
\nonumber \\
& =2%
\genfrac{}{}{0pt}{}{:}{:}%
\exp \left[  -2\left(  \alpha-a\right)  \left(  \alpha^{\ast}-a^{\dagger
}\right)  \right]
\genfrac{}{}{0pt}{}{:}{:}%
\nonumber \\
& =2%
\genfrac{}{}{0pt}{}{:}{:}%
\exp \left[  -\left(  p-P\right)  ^{2}-\left(  q-Q\right)  ^{2}\right]
\genfrac{}{}{0pt}{}{:}{:}%
,\label{22}%
\end{align}
thus the Weyl ordered form of pure coherent state $\left \vert \alpha
\right \rangle \left \langle \alpha \right \vert $ is a Gaussian in $p-q$ space.
Combining Eqs. (21), (8) and (20) yields
\begin{align}
2\pi \mathtt{Tr}\left[  \rho \Delta \left(  \alpha,\alpha^{\ast}\right)  \right]
& =4\int \mathtt{d}^{2}\beta \left \langle -\beta \right \vert \rho \left \vert
\beta \right \rangle \mathtt{Tr}\left \{
\genfrac{}{}{0pt}{}{:}{:}%
\exp \left[  2\left(  \beta^{\ast}a-a^{\dagger}\beta+a^{\dagger}a\right)
\right]
\genfrac{}{}{0pt}{}{:}{:}%
\Delta \left(  \alpha,\alpha^{\ast}\right)  \right \} \nonumber \\
& =4\int \mathtt{d}^{2}\beta \left \langle -\beta \right \vert \rho \left \vert
\beta \right \rangle \mathtt{Tr}\left[  2\int \mathtt{d}^{2}\alpha^{\prime}%
\exp \left[  2\left(  \beta^{\ast}\alpha^{\prime}-\alpha^{\prime \ast}%
\beta+\alpha^{\prime \ast}\alpha^{\prime}\right)  \right]  \Delta \left(
\alpha^{\prime},\alpha^{\prime \ast}\right)  \Delta \left(  \alpha,\alpha^{\ast
}\right)  \right] \nonumber \\
& =2\int \frac{\mathtt{d}^{2}\beta}{\pi}\left \langle -\beta \right \vert
\rho \left \vert \beta \right \rangle \int \mathtt{d}^{2}\alpha^{\prime}\exp \left[
2\left(  \beta^{\ast}\alpha^{\prime}-\alpha^{\prime \ast}\beta+\alpha
^{\prime \ast}\alpha \right)  \right]  \delta \left(  \alpha-\alpha^{\prime
}\right)  \delta \left(  \alpha^{\ast}-\alpha^{\prime \ast}\right) \nonumber \\
& =2\int \frac{\mathtt{d}^{2}\beta}{\pi}\left \langle -\beta \right \vert
\rho \left \vert \beta \right \rangle \exp \left[  2\left(  \beta^{\ast}%
\alpha-\alpha^{\ast}\beta+\alpha^{\ast}\alpha \right)  \right]  ,\label{23}%
\end{align}
which is just an alternate expression of the Wigner function of $\rho,$
comparing (21) with (23) we see that the latter is just the result of
replacing $a\rightarrow \alpha,$ $a^{\dagger}\rightarrow \alpha^{\ast},$ in the
former, this is because that the right hand side of (21) is in Weyl ordering.

Now we examine what is the Weyl ordering of $\delta \left(  p-P\right)
\delta \left(  q-Q\right)  .$ Using the completeness relation of $\left \vert
q\right \rangle ,$ the coordinate eigenstate, and the completeness relation of
the momentum eigenstate $\left \vert p\right \rangle ,$ $\left \langle q\right.
\left \vert p\right \rangle =\frac{1}{\sqrt{2\pi}}e^{\mathtt{i}pq},$ we have
\begin{align}
\delta \left(  p-P\right)  \delta \left(  q-Q\right)   & =\int \mathtt{d}%
p^{\prime}\left \vert p^{\prime}\right \rangle \left \langle p^{\prime
}\right \vert \delta \left(  p-P\right)  \delta \left(  q-Q\right)
\int \mathtt{d}q^{\prime}\left \vert q^{\prime}\right \rangle \left \langle
q^{\prime}\right \vert \nonumber \\
& =\frac{1}{\sqrt{2\pi}}\int \mathtt{d}p^{\prime}\left \vert p^{\prime
}\right \rangle \int \mathtt{d}q^{\prime}\left \langle q^{\prime}\right \vert
\delta \left(  p-p^{\prime}\right)  \delta \left(  q-q^{\prime}\right)
e^{-\mathtt{i}p^{\prime}q^{\prime}}\nonumber \\
& =\frac{1}{\sqrt{2\pi}}\left \vert p\right \rangle \left \langle q\right \vert
e^{-\mathtt{i}pq}.\label{24}%
\end{align}
The overlap between $\left \langle q\right \vert $ and the coherent state is%
\begin{equation}
\left \langle q\right \vert \left.  \beta \right \rangle =\pi^{-1/4}\exp \left \{
-\frac{q^{2}}{2}+\sqrt{2}q\beta-\frac{1}{2}\beta^{2}-\frac{1}{2}|\beta
|^{2}\right \}  ,\label{25}%
\end{equation}
and%
\begin{equation}
\left \langle -\beta \right.  \left \vert p\right \rangle =\pi^{-1/4}\exp \left \{
-\frac{p^{2}}{2}-\sqrt{2}ip\beta^{\ast}+\frac{1}{2}\beta^{\ast2}-\frac{1}%
{2}|\beta|^{2}\right \}  .\label{26}%
\end{equation}
Substituting (24) into (21) and using (25)-(26) lead to%
\begin{align}
& \delta \left(  p-P\right)  \delta \left(  q-Q\right) \nonumber \\
& =\frac{\sqrt{2}}{\pi}\int \frac{d^{2}\beta}{\pi}%
\genfrac{}{}{0pt}{}{:}{:}%
\left \langle -\beta \right \vert \left.  p\right \rangle \left \langle
q\right \vert e^{-\mathtt{i}pq}\left \vert \beta \right \rangle \exp \left[
2\left(  \beta^{\ast}a-a^{\dagger}\beta+a^{\dagger}a\right)  \right]
\genfrac{}{}{0pt}{}{:}{:}%
\nonumber \\
& =\frac{\sqrt{2}}{\pi}e^{-\frac{q^{2}+p^{2}}{2}-\mathtt{i}pq}\int
\frac{\mathtt{d}^{2}\beta}{\pi}%
\genfrac{}{}{0pt}{}{:}{:}%
\exp \left \{  -|\beta|^{2}+\sqrt{2}q\beta-\sqrt{2}\mathtt{i}p\beta^{\ast
}\right \} \nonumber \\
& \times \exp \left[  2\left(  \beta^{\ast}a-a^{\dagger}\beta+a^{\dagger
}a\right)  -\frac{\beta^{2}}{2}+\frac{\beta^{\ast2}}{2}\right]
\genfrac{}{}{0pt}{}{:}{:}%
\nonumber \\
& =\frac{1}{\pi}%
\genfrac{}{}{0pt}{}{:}{:}%
\exp \{ \sqrt{2}q\left(  a-a^{\dagger}\right)  +\sqrt{2}\mathtt{i}p\left(
a+a^{\dagger}\right)  -2\mathtt{i}pq+a^{\dagger2}-a^{2}-a^{\dagger}a\}%
\genfrac{}{}{0pt}{}{:}{:}%
\nonumber \\
& =\frac{1}{\pi}%
\genfrac{}{}{0pt}{}{:}{:}%
\exp[-2\mathtt{i}\left(  q-Q\right)  \left(  p-P\right)  ]%
\genfrac{}{}{0pt}{}{:}{:}%
.\label{27}%
\end{align}
Similarly, we can derive
\begin{align}
\delta \left(  q-Q\right)  \delta \left(  p-P\right)   & =2\int \frac
{\mathtt{d}^{2}\beta}{\pi}%
\genfrac{}{}{0pt}{}{:}{:}%
\left \langle -\beta \right \vert \left.  q\right \rangle \left \langle
p\right \vert e^{\mathtt{i}pq}\left \vert \beta \right \rangle \exp \left[
2\left(  \beta^{\ast}a-a^{\dagger}\beta+a^{\dagger}a\right)  \right]
\genfrac{}{}{0pt}{}{:}{:}%
\nonumber \\
& =\frac{1}{\pi}%
\genfrac{}{}{0pt}{}{:}{:}%
\exp[2\mathtt{i}\left(  q-Q\right)  \left(  p-P\right)  ]%
\genfrac{}{}{0pt}{}{:}{:}%
.\label{28}%
\end{align}
Eqs. (27)-(28) are the Weyl ordered forms of $\delta \left(  p-P\right)
\delta \left(  q-Q\right)  $ and $\delta \left(  q-Q\right)  \delta \left(
p-P\right)  ,$ respectively.

\section{The new transformation of Wigner operator}

Taking $\frac{1}{\pi}%
\genfrac{}{}{0pt}{}{:}{:}%
\exp[-2\mathtt{i}\left(  q-Q\right)  \left(  p-P\right)  ]%
\genfrac{}{}{0pt}{}{:}{:}%
$as an integration kernel of the following integration transformation with the
result $%
\genfrac{}{}{0pt}{}{:}{:}%
K\left(  P,Q\right)
\genfrac{}{}{0pt}{}{:}{:}%
,$
\begin{equation}
\iint_{-\infty}^{\infty}\frac{\mathtt{d}p\mathtt{d}q}{\pi}f\left(  p,q\right)
%
\genfrac{}{}{0pt}{}{:}{:}%
\exp[-2\mathtt{i}\left(  q-Q\right)  \left(  p-P\right)  ]%
\genfrac{}{}{0pt}{}{:}{:}%
=%
\genfrac{}{}{0pt}{}{:}{:}%
K\left(  P,Q\right)
\genfrac{}{}{0pt}{}{:}{:}%
,\label{29}%
\end{equation}
then from (27) we have%
\begin{equation}%
\genfrac{}{}{0pt}{}{:}{:}%
K\left(  P,Q\right)
\genfrac{}{}{0pt}{}{:}{:}%
=\iint_{-\infty}^{\infty}\mathtt{d}p\mathtt{d}qf\left(  p,q\right)
\delta \left(  p-P\right)  \delta \left(  q-Q\right)  =f\left(  p,q\right)
|_{p\rightarrow P,\text{ }q\rightarrow Q,\text{ }P\text{ before }Q},\label{30}%
\end{equation}
this is the integration formula for quantizing classical function $f(p,q)$ as
$P-Q$ ordering of operators. On the other hand, from (28) we have
\begin{align}
& \iint_{-\infty}^{\infty}\frac{\mathtt{d}p\mathtt{d}q}{\pi}f(p,q)%
\genfrac{}{}{0pt}{}{:}{:}%
\exp[2\mathtt{i}\left(  q-Q\right)  \left(  p-P\right)  ]%
\genfrac{}{}{0pt}{}{:}{:}%
\nonumber \\
& =\iint_{-\infty}^{\infty}\mathtt{d}p\mathtt{d}qf(p,q)\delta \left(
q-Q\right)  \delta \left(  p-P\right)  =f\left(  p,q\right)  |_{q\rightarrow
Q,\text{ }p\rightarrow P,\text{ }Q\text{ before }P},\label{31}%
\end{align}
this is the scheme of quantizing classical function $f(p,q)$ as $Q-P$ ordering
of operators.

By noticing (5) we see%

\begin{align}
& \frac{1}{\pi}%
\genfrac{}{}{0pt}{}{:}{:}%
\exp[-2\mathtt{i}\left(  q-Q\right)  \left(  p-P\right)  ]%
\genfrac{}{}{0pt}{}{:}{:}%
\nonumber \\
& =\frac{1}{\pi}\iint \mathtt{d}p^{\prime}\mathtt{d}q^{\prime}e^{-2\mathtt{i}%
\left(  q-q^{\prime}\right)  \left(  p-p^{\prime}\right)  }%
\genfrac{}{}{0pt}{}{:}{:}%
\delta \left(  q^{\prime}-Q\right)  \delta \left(  p^{\prime}-P\right)
\genfrac{}{}{0pt}{}{:}{:}%
\nonumber \\
& =\frac{1}{\pi}\iint \mathtt{d}p^{\prime}\mathtt{d}q^{\prime}\Delta \left(
q^{\prime},p^{\prime}\right)  e^{-2\mathtt{i}\left(  p-p^{\prime}\right)
\left(  q-q^{\prime}\right)  }.\label{32}%
\end{align}
It then follows from (32) and (27) that
\begin{equation}
\frac{1}{\pi}\iint \mathtt{d}p^{\prime}\mathtt{d}q^{\prime}\Delta \left(
q^{\prime},p^{\prime}\right)  e^{-2\mathtt{i}\left(  p-p^{\prime}\right)
\left(  q-q^{\prime}\right)  }=\delta \left(  p-P\right)  \delta \left(
q-Q\right)  .\label{33}%
\end{equation}
Similarly we can derive%
\begin{equation}
\frac{1}{\pi}\iint \mathtt{d}p^{\prime}\mathtt{d}q^{\prime}\Delta \left(
q^{\prime},p^{\prime}\right)  e^{2\mathtt{i}\left(  p-p^{\prime}\right)
\left(  q-q^{\prime}\right)  }=\delta \left(  q-Q\right)  \delta \left(
p-P\right)  ,\label{34}%
\end{equation}
so $e^{\pm2\mathtt{i}\left(  p-p^{\prime}\right)  \left(  q-q^{\prime}\right)
}/\pi$ can be considered the classical Weyl correspondence of $\delta \left(
q-Q\right)  \delta \left(  p-P\right)  $ and $\delta \left(  p-P\right)
\delta \left(  q-Q\right)  ,$ respectively$.$ Moreover, the inverse transform
of (32) is%
\begin{align}
& \iint \frac{\mathtt{d}q\mathtt{d}p}{\pi}%
\genfrac{}{}{0pt}{}{:}{:}%
\exp[-2\mathtt{i}\left(  q-Q\right)  \left(  p-P\right)  ]%
\genfrac{}{}{0pt}{}{:}{:}%
e^{2\mathtt{i}\left(  p-p^{\prime}\right)  \left(  q-q^{\prime}\right)
}\nonumber \\
& =\iint \frac{\mathtt{d}q\mathtt{d}p}{\pi}\iint dp^{\prime \prime}%
dq^{\prime \prime}\Delta \left(  q^{\prime \prime},p^{\prime \prime}\right)
e^{-2\mathtt{i}\left(  p-p^{\prime \prime}\right)  \left(  q-q^{\prime \prime
}\right)  +2\mathtt{i}\left(  p-p^{\prime}\right)  \left(  q-q^{\prime
}\right)  }\nonumber \\
& =\iint dp^{\prime \prime}dq^{\prime \prime}\Delta \left(  q^{\prime \prime
},p^{\prime \prime}\right)  e^{-2i\left(  p^{\prime \prime}q^{\prime \prime
}-p^{\prime}q^{\prime}\right)  }\delta \left(  q^{\prime}-q^{\prime \prime
}\right)  \delta \left(  p^{\prime}-p^{\prime \prime}\right)  =\Delta \left(
q^{\prime},p^{\prime}\right)  .\label{35}%
\end{align}
which means%
\begin{equation}
\iint \mathtt{d}q\mathtt{d}p\delta \left(  p-P\right)  \delta \left(  q-Q\right)
e^{2\mathtt{i}\left(  p-p^{\prime}\right)  \left(  q-q^{\prime}\right)
}=\Delta \left(  q^{\prime},p^{\prime}\right)  ,\label{36}%
\end{equation}
or%
\begin{equation}
\iint \mathtt{d}q\mathtt{d}p\delta \left(  q-Q\right)  \delta \left(  p-P\right)
e^{-2\mathtt{i}\left(  p-p^{\prime}\right)  \left(  q-q^{\prime}\right)
}=\Delta \left(  q^{\prime},p^{\prime}\right)  .\label{37}%
\end{equation}
Eqs. (33)-(37) are new transformations of the Wigner operator in $q-p$ phase space.

\section{The new transformation in phase space}

Further, multiplying both sides of (35) from the left by $\iint \mathtt{d}%
q^{\prime}\mathtt{d}p^{\prime}h\left(  p^{\prime},q^{\prime}\right)  $ we
obtain
\begin{align}
& \iint \mathtt{d}q^{\prime}\mathtt{d}p^{\prime}h\left(  p^{\prime},q^{\prime
}\right)  \Delta \left(  q^{\prime},p^{\prime}\right) \nonumber \\
& =\iint \mathtt{d}q^{\prime}\mathtt{d}p^{\prime}h\left(  p^{\prime},q^{\prime
}\right)  \iint \frac{\mathtt{d}q\mathtt{d}p}{\pi}%
\genfrac{}{}{0pt}{}{:}{:}%
\exp[-2i\left(  q-Q\right)  \left(  p-P\right)  ]%
\genfrac{}{}{0pt}{}{:}{:}%
e^{2\mathtt{i}\left(  p-p^{\prime}\right)  \left(  q-q^{\prime}\right)
}\nonumber \\
& =\iint \frac{\mathtt{d}q\mathtt{d}p}{\pi}%
\genfrac{}{}{0pt}{}{:}{:}%
\exp[-2\mathtt{i}\left(  q-Q\right)  \left(  p-P\right)  ]%
\genfrac{}{}{0pt}{}{:}{:}%
G\left(  p,q\right)  ,\label{38}%
\end{align}
where we have introduced
\begin{equation}
G\left(  p,q\right)  \equiv \frac{1}{\pi}\iint \mathtt{d}q^{\prime}%
\mathtt{d}p^{\prime}h\left(  p^{\prime},q^{\prime}\right)  e^{2\mathtt{i}%
\left(  p-p^{\prime}\right)  \left(  q-q^{\prime}\right)  },\label{39}%
\end{equation}
this is a new interesting transformation, because when $h\left(  p^{\prime
},q^{\prime}\right)  =1,$
\begin{equation}
\frac{1}{\pi}\iint \mathtt{d}q^{\prime}\mathtt{d}p^{\prime}e^{2\mathtt{i}%
\left(  p-p^{\prime}\right)  \left(  q-q^{\prime}\right)  }=\int_{-\infty
}^{\infty}\mathtt{d}q^{\prime}\delta \left(  q-q^{\prime}\right)
e^{2\mathtt{i}p\left(  q-q^{\prime}\right)  }=1.\label{40}%
\end{equation}
The inverse of (39) is%
\begin{equation}
\iint \frac{dqdp}{\pi}e^{-2i\left(  p-p^{\prime}\right)  \left(  q-q^{\prime
}\right)  }G\left(  p,q\right)  =h\left(  p^{\prime},q^{\prime}\right)
.\label{41}%
\end{equation}
In fact, substituting (39) into the the left-hand side of (41) yields%
\begin{align}
& \iint_{-\infty}^{\infty}\frac{\mathtt{d}q\mathtt{d}p}{\pi}\iint
\frac{\mathtt{d}q^{\prime \prime}\mathtt{d}p^{\prime \prime}}{\pi}%
h(p^{\prime \prime},q^{\prime \prime})e^{2\mathtt{i}\left[  \left(
p-p^{\prime \prime}\right)  \left(  q-q^{\prime \prime}\right)  -\left(
p-p^{\prime}\right)  \left(  q-q^{\prime}\right)  \right]  }\nonumber \\
& =\iint_{-\infty}^{\infty}\mathtt{d}q^{\prime \prime}\mathtt{d}p^{\prime
\prime}h(p^{\prime \prime},q^{\prime \prime})e^{2\mathtt{i}\left(
p^{\prime \prime}q^{\prime \prime}-p^{\prime}q^{\prime}\right)  }\delta \left(
p^{\prime \prime}-p^{\prime}\right)  \delta \left(  q^{\prime \prime}-q^{\prime
}\right)  =h(p^{\prime},q^{\prime}).\label{42}%
\end{align}
This transformation's Parsval-like theorem is
\begin{align}
& \iint_{-\infty}^{\infty}\frac{\mathtt{d}q\mathtt{d}p}{\pi}|h(p,q)|^{2}%
\nonumber \\
& =\iint \frac{\mathtt{d}q^{\prime}\mathtt{d}p^{\prime}}{\pi}|G\left(
p^{\prime},q^{\prime}\right)  |^{2}\iint \frac{\mathtt{d}p^{\prime \prime
}\mathtt{d}q^{\prime \prime}}{\pi}e^{2i\left(  p^{\prime \prime}q^{\prime \prime
}-p^{\prime}q^{\prime}\right)  }\iint_{-\infty}^{\infty}\frac{\mathtt{d}%
q\mathtt{d}p}{\pi}e^{2i\left[  \left(  -p^{\prime \prime}p-q^{\prime \prime
}q\right)  +\left(  pp^{\prime}+q^{\prime}q\right)  \right]  }\nonumber \\
& =\iint \frac{\mathtt{d}q^{\prime}\mathtt{d}p^{\prime}}{\pi}|G\left(
p^{\prime},q^{\prime}\right)  |^{2}\iint \mathtt{d}p^{\prime \prime}%
\mathtt{d}q^{\prime \prime}e^{2i\left(  p^{\prime \prime}q^{\prime \prime
}-p^{\prime}q^{\prime}\right)  }\delta \left(  q^{\prime}-q^{\prime \prime
}\right)  \delta \left(  p^{\prime}-p^{\prime \prime}\right)  =\iint
\frac{\mathtt{d}q^{\prime}\mathtt{d}p^{\prime}}{\pi}|G\left(  p^{\prime
},q^{\prime}\right)  |^{2}.\label{43}%
\end{align}

\section{P-Q ordering and Q-P ordering to Weyl ordering}

We now use the above transformation to discuss some operator ordering
problems. For instance, from the integration formula%

\begin{equation}
\iint \limits_{-\infty}^{\infty}\frac{\mathtt{d}x\mathtt{d}y}{\pi}x^{m}%
y^{r}\exp[2\mathtt{i}\left(  y-s\right)  \left(  x-t\right)  ]=\left(
\frac{1}{\sqrt{2}}\right)  ^{m+r}\left(  -\mathtt{i}\right)  ^{r}%
H_{m,r}\left(  \sqrt{2}t,\mathtt{i}\sqrt{2}s\right)  ,\label{44}%
\end{equation}
where $H_{m,r\text{ }}$is the two-variable Hermite polynomials [12-13],%
\begin{equation}
H_{m,r}(t,s)=\sum_{l=0}^{\min(m,r)}\frac{m!r!(-1)^{l}}{l!(m-l)!(r-l)!}%
t^{m-l}s^{r-l}.\label{45}%
\end{equation}
Eq. (44) can be proved as follows:
\begin{align}
\text{L.H.S. of (44)}  & =e^{2\mathtt{i}st}\left(  \frac{\partial}{\partial
t}\right)  ^{r}\left(  \frac{\partial}{\partial s}\right)  ^{m}\iint
\limits_{-\infty}^{\infty}\frac{\mathtt{d}x\mathtt{d}y}{\pi}e^{2\mathtt{i}%
xy}\exp[-2\mathtt{i}yt-2\mathtt{i}sx]\nonumber \\
& =e^{2\mathtt{i}st}\left(  \frac{\partial}{\partial t}\right)  ^{r}\left(
\frac{\partial}{\partial s}\right)  ^{m}\int_{-\infty}^{\infty}\mathtt{d}%
xe^{-2\mathtt{i}sx}\delta \left(  x-t\right) \nonumber \\
& =e^{2\mathtt{i}st}\left(  \frac{\partial}{\partial t}\right)  ^{r}\left(
\frac{\partial}{\partial s}\right)  ^{m}e^{-2\mathtt{i}st}=\text{R.H.S. of
(44).}\label{46}%
\end{align}
Using (28) and (44) we know%
\begin{align}
Q^{m}P^{r}  & =\iint_{-\infty}^{\infty}\mathtt{d}p\mathtt{d}qq^{m}p^{r}%
\delta \left(  q-Q\right)  \delta \left(  p-P\right) \nonumber \\
& =\iint_{-\infty}^{\infty}\frac{\mathtt{d}p\mathtt{d}q}{\pi}q^{m}p^{r}%
\genfrac{}{}{0pt}{}{:}{:}%
\exp[2\mathtt{i}\left(  p-P\right)  \left(  q-Q\right)  ]%
\genfrac{}{}{0pt}{}{:}{:}%
\nonumber \\
& =\left(  \frac{1}{\sqrt{2}}\right)  ^{m+r}\left(  -\mathtt{i}\right)  ^{r}%
\genfrac{}{}{0pt}{}{:}{:}%
H_{m,r}\left(  \sqrt{2}Q,\mathtt{i}\sqrt{2}P\right)
\genfrac{}{}{0pt}{}{:}{:}%
,\label{47}%
\end{align}
this is a simpler way to put $Q^{m}P^{r}$ into its Weyl ordering. Similarly,
using (27) and the complex conjugate of (44) we see that the Weyl ordered form
of $P^{r}Q^{m}$ is
\begin{align}
P^{r}Q^{m}  & =\iint_{-\infty}^{\infty}\mathtt{d}p\mathtt{d}qp^{r}q^{m}%
\delta \left(  p-P\right)  \delta \left(  q-Q\right) \nonumber \\
& =\iint_{-\infty}^{\infty}\frac{\mathtt{d}p\mathtt{d}q}{\pi}%
\genfrac{}{}{0pt}{}{:}{:}%
\exp[-2\mathtt{i}\left(  q-Q\right)  \left(  p-P\right)  ]%
\genfrac{}{}{0pt}{}{:}{:}%
q^{m}p^{r}\nonumber \\
& =\left(  \frac{1}{\sqrt{2}}\right)  ^{m+r}\left(  \mathtt{i}\right)  ^{r}%
\genfrac{}{}{0pt}{}{:}{:}%
H_{m,r}\left(  \sqrt{2}Q,-\mathtt{i}\sqrt{2}P\right)
\genfrac{}{}{0pt}{}{:}{:}%
.\label{48}%
\end{align}

\section{Weyl ordering to P-Q ordering and Q-P ordering}

According to (39) and (41) we know that the inverse transform of (44) is
\begin{equation}
\iint \frac{\mathtt{d}s\mathtt{d}t}{\pi}\left(  \frac{1}{\sqrt{2}}\right)
^{m+r}\left(  -\mathtt{i}\right)  ^{r}H_{m,r}\left(  \sqrt{2}t,\mathtt{i}%
\sqrt{2}s\right)  e^{-2\mathtt{i}\left(  y-s\right)  \left(  x-t\right)
}=x^{m}y^{r},\label{49}%
\end{equation}
which is a new integration formula. Then from (27) and (49) we have%
\begin{align}
& \left(  \frac{1}{\sqrt{2}}\right)  ^{m+r}\left(  -\mathtt{i}\right)
^{r}H_{m,r}\left(  \sqrt{2}Q,\mathtt{i}\sqrt{2}P\right)  |_{P\text{ before }%
Q}\nonumber \\
& =\left(  \frac{1}{\sqrt{2}}\right)  ^{m+r}\left(  -\mathtt{i}\right)
^{r}\iint \mathtt{d}p\mathtt{d}q\delta \left(  p-P\right)  \delta \left(
q-Q\right)  H_{m,r}\left(  \sqrt{2}q,\mathtt{i}\sqrt{2}p\right) \nonumber \\
& =\left(  \frac{1}{\sqrt{2}}\right)  ^{m+r}\left(  -\mathtt{i}\right)
^{r}\iint \frac{\mathtt{d}p\mathtt{d}q}{\pi}H_{m,r}\left(  \sqrt{2}%
q,\mathtt{i}\sqrt{2}p\right)
\genfrac{}{}{0pt}{}{:}{:}%
e^{-2\mathtt{i}\left(  q-Q\right)  \left(  p-P\right)  }%
\genfrac{}{}{0pt}{}{:}{:}%
=%
\genfrac{}{}{0pt}{}{:}{:}%
Q^{m}P^{r}%
\genfrac{}{}{0pt}{}{:}{:}%
.\label{50}%
\end{align}
Due to (45) we see%
\begin{equation}
\left(  \frac{1}{\sqrt{2}}\right)  ^{m+r}\left(  -\mathtt{i}\right)
^{r}H_{m,r}\left(  \sqrt{2}Q,\mathtt{i}\sqrt{2}P\right)  |_{P\text{ before }%
Q}=\sum_{l=0}\left(  \frac{\mathtt{i}}{2}\right)  ^{l}l!\binom{r}{l}\binom
{m}{l}P^{r-l}Q^{m-l},\label{51}%
\end{equation}
so (50)-(51) leads to%
\begin{equation}%
\genfrac{}{}{0pt}{}{:}{:}%
Q^{m}P^{r}%
\genfrac{}{}{0pt}{}{:}{:}%
=\sum_{l=0}\left(  \frac{\mathtt{i}}{2}\right)  ^{l}l!\binom{r}{l}\binom{m}%
{l}P^{r-l}Q^{m-l},\label{52}%
\end{equation}
Eq. (50) or Eq. (52) is the fundamental formula of converting Weyl ordered
operator to its $P-Q$ ordering.

Similarly, from (28) and the hermite conjugate of (49) we have
\begin{align}
& \left(  \frac{1}{\sqrt{2}}\right)  ^{m+r}\left(  \mathtt{i}\right)
^{r}H_{m,r}\left(  \sqrt{2}Q,-\mathtt{i}\sqrt{2}P\right)  |_{Q\text{ before
}P\text{ }}\nonumber \\
& =\iint \mathtt{d}p\mathtt{d}q\delta \left(  q-Q\right)  \delta \left(
p-P\right)  \left(  \frac{1}{\sqrt{2}}\right)  ^{m+r}\left(  \mathtt{i}%
\right)  ^{r}H_{m,r}\left(  \sqrt{2}q,-\mathtt{i}\sqrt{2}p\right) \nonumber \\
& =\iint \frac{\mathtt{d}p\mathtt{d}q}{\pi}\left(  \frac{1}{\sqrt{2}}\right)
^{m+r}\left(  \mathtt{i}\right)  ^{r}H_{m,r}\left(  \sqrt{2}q,-\mathtt{i}%
\sqrt{2}p\right)
\genfrac{}{}{0pt}{}{:}{:}%
e^{2\mathtt{i}\left(  q-Q\right)  \left(  p-P\right)  }%
\genfrac{}{}{0pt}{}{:}{:}%
\nonumber \\
& =%
\genfrac{}{}{0pt}{}{:}{:}%
Q^{m}P^{r}%
\genfrac{}{}{0pt}{}{:}{:}%
=%
\genfrac{}{}{0pt}{}{:}{:}%
P^{r}Q^{m}%
\genfrac{}{}{0pt}{}{:}{:}%
,\label{53}%
\end{align}
so%
\begin{equation}%
\genfrac{}{}{0pt}{}{:}{:}%
Q^{m}P^{r}%
\genfrac{}{}{0pt}{}{:}{:}%
=\sum_{l=0}\left(  \frac{-\mathtt{i}}{2}\right)  ^{l}l!\binom{r}{l}\binom
{m}{l}Q^{m-l}P^{r-l},\label{54}%
\end{equation}
this is the fundamental formula of converting Weyl ordered operator to its
$Q-P$ ordering, which is in contrast to (52).

\section{Q-P ordering to P-Q ordering and vice versa}

Combining (47) and (52) together we derive
\begin{align}
Q^{m}P^{r}  & =\sum_{l=0}\frac{m!r!}{l!(m-l)!(r-l)!}(\frac{\mathtt{i}}{2})^{l}%
\genfrac{}{}{0pt}{}{:}{:}%
Q^{m-l}P^{r-l}%
\genfrac{}{}{0pt}{}{:}{:}%
\nonumber \\
& =\sum_{l=0}\frac{m!r!}{l!(m-l)!(r-l)!}(\frac{\mathtt{i}}{2})^{l}\sum
_{k=0}\left(  \frac{\mathtt{i}}{2}\right)  ^{k}k!\binom{r-l}{k}\binom{m-l}%
{k}P^{r-l-k}Q^{m-l-k}\nonumber \\
& =\sum_{l=0}\sum_{k=0}\frac{m!r!}{l!(m-l-k)!(r-l-k)!k!}(\frac{\mathtt{i}}%
{2})^{l+k}P^{r-l-k}Q^{m-l-k}\nonumber \\
& =\sum_{k=0}\frac{m!r!}{(m-k)!(r-k)!k!}(\mathtt{i})^{k}P^{r-k}Q^{m-k}%
,\label{55}%
\end{align}
which puts $Q^{m}P^{r}$ to its $P-Q$ ordering. It then follows the commutator%
\begin{equation}
\left[  Q^{m},P^{r}\right]  =\sum_{k=1}\frac{m!r!}{(m-k)!(r-k)!k!}%
(\mathtt{i})^{k}P^{r-k}Q^{m-k}.\label{56}%
\end{equation}
On the other hand, from (48), (45) and (54) we have%
\begin{align}
P^{r}Q^{m}  & =\left(  \frac{1}{\sqrt{2}}\right)  ^{m+r}\left(  \mathtt{i}%
\right)  ^{r}%
\genfrac{}{}{0pt}{}{:}{:}%
H_{m,r}\left(  \sqrt{2}Q,-\mathtt{i}\sqrt{2}P\right)
\genfrac{}{}{0pt}{}{:}{:}%
\nonumber \\
& =%
\genfrac{}{}{0pt}{}{:}{:}%
\sum_{l=0}\frac{m!r!}{l!(m-l)!(r-l)!}(\frac{-\mathtt{i}}{2})^{l}%
\genfrac{}{}{0pt}{}{:}{:}%
Q^{m-l}P^{r-l}%
\genfrac{}{}{0pt}{}{:}{:}%
\nonumber \\
& =\sum_{l=0}\frac{m!r!}{l!(m-l)!(r-l)!}(\frac{-\mathtt{i}}{2})^{l}\sum
_{k=0}\left(  \frac{-\mathtt{i}}{2}\right)  ^{k}k!\binom{r-l}{k}\binom{m-l}%
{k}Q^{m-l-k}P^{r-l-k}\nonumber \\
& =\sum_{k=0}\frac{m!r!}{(m-k)!(r-k)!k!}(-\mathtt{i})^{k}Q^{m-k}%
P^{r-k},\label{57}%
\end{align}
which puts $P^{r}Q^{m}$ to its $Q-P$ ordering. Thus (56) is also equal to
\begin{equation}
\left[  Q^{m},P^{r}\right]  =\sum_{k=1}\frac{m!r!}{(m-k)!(r-k)!k!}%
(-\mathtt{i})^{k}Q^{m-k}P^{r-k}.\label{58}%
\end{equation}

\section{$P-Q$ ordering or $Q-P$ ordering expansion of $\left(  P+Q\right)
^{n}$}

Due to%
\begin{align}
\left(  P+Q\right)  ^{n}  & =\frac{\mathtt{d}^{n}}{\mathtt{d}\lambda^{n}%
}\left.  e^{\lambda \left(  P+Q\right)  }\right \vert _{\lambda=0}%
=\frac{\mathtt{d}^{n}}{\mathtt{d}\lambda^{n}}\left.
\genfrac{}{}{0pt}{}{:}{:}%
e^{\lambda \left(  P+Q\right)  }%
\genfrac{}{}{0pt}{}{:}{:}%
\right \vert _{\lambda=0}\nonumber \\
& =%
\genfrac{}{}{0pt}{}{:}{:}%
\left(  P+Q\right)  ^{n}%
\genfrac{}{}{0pt}{}{:}{:}%
=\sum_{l=0}^{n}\binom{n}{l}%
\genfrac{}{}{0pt}{}{:}{:}%
Q^{l}P^{n-l}%
\genfrac{}{}{0pt}{}{:}{:}%
,\label{59}%
\end{align}
substituting (52) into (59) we derive%
\begin{equation}
\left(  P+Q\right)  ^{n}=\sum_{l=0}^{n}\binom{n}{l}\sum_{k=0}\left(
\frac{\mathtt{i}}{2}\right)  ^{k}k!\binom{l}{k}\binom{n-l}{k}P^{l-k}%
Q^{n-l-k},\label{60}%
\end{equation}
or using (54) we have%
\begin{equation}
\left(  P+Q\right)  ^{n}=\sum_{l=0}^{n}\binom{n}{l}\sum_{k=0}\left(
\frac{-\mathtt{i}}{2}\right)  ^{k}k!\binom{l}{k}\binom{n-l}{k}Q^{l-k}%
P^{n-l-k}.\label{61}%
\end{equation}

In sum, by virtue of the formula of operators' Weyl ordering expansion and the
technique of integration within Weyl ordered product of operators we have
found new two-fold integration transformation about the Wigner operator
$\Delta \left(  q^{\prime},p^{\prime}\right)  $ in phase space quantum
mechanics, which provides us with a new approach for deriving mutual
converting formulas among $Q-P$ ordering, $P-Q$ ordering and Weyl ordering of
operators. A new $c$-number two-fold integration transformation in $p-q$ phase
space (Eq. (39)-(41)) is also proposed, we expect that it may have other uses
in theoretical physics. In this way, the contents of phase space quantum
mechanics [14] can be enriched.


\begin{thebibliography}{99}                                                                                               %
\bibitem {r1}H. Z. Weyl, Physics, 46 (1927) 1

\bibitem {r2}E. Wigner, Phys. Rev. 40 (1932) 749; G. S. Agarwal and E. Wolf,
Phys. Rev. D \textbf{2 (}1970) 2161; M. Hillery, R. Connel, M. Scully and E.
Wigner, Phys. Rep. 106 (1984) 121;V. Bu\v{z}ek, C. H. Keitel and P. L. Knight,
Phys. Rev. A \textbf{51} (1995) 2575; H. Moyal, Proc. Camb. Phil. Soc. 45
(1949) 99

\bibitem {r3}H. Lee, Phys. Rep. 259 (1995) 150

\bibitem {r4}N. L. Balazs and B. K. Jennings, Phys. Rep. 104 (1984) 347; C.
Zachos, Inter. J. Mod. Phys. A 17, (2002) 297

\bibitem {r5}W. Schleich, Quantum Optics in Phase Space, Wiley-VCH, Berlin
2001 and many references therein

\bibitem {r6}Hong-yi Fan,\emph{\ }J. Phys. A 25 (1992) 3443

\bibitem {r7}Hong-yi Fan, Ann. Phys. 323 (2008) 500

\bibitem {r8}A. W\"{u}nsche, J. Opt. B: Quantum Semiclass. Opt. 1 (1999) R11;
Hong-yi Fan, J. Opt. B: Quantum Semiclass. Opt. 5 (2003) R147

\bibitem {r9}Hong-yi Fan and H. R. Zaidi, Phys. Lett. A \textbf{123} (1987)
303; Hong-yi Fan and Tu-nan Ruan, Commun. Theor. Phys. \textbf{2} (1983) 1563;
\textbf{3} (1984) 345

\bibitem {r10}J. R. Klauder and B. -S. Skagerstam, Coherent States, World
Scientific, Singapore, 1985

\bibitem {r11}R. J. Glauber, Phys. Rev. 130 (1963) 2529; 131 (1963) 2766

\bibitem {r12}A. Erd\`{e}lyi, Higher Transcendental Functions, The Bateman
Manuscript Project, McGraw Hill, 1953

\bibitem {r13}Hong-yi Fan and Jun-hua Chen, Phys. Lett. A 303 (2002) 311

\bibitem {r14}K. Vogel and H. Risken, Phys. Rev. A 40, (1989) 2847; P.
Kasperkovitz and M. Peev, Ann. Phys. 230 (1994) 21
\end{thebibliography}
\end{document}